\pdfoutput=1

\documentclass[11pt]{article}

\usepackage[preprint]{acl}

\usepackage{times}
\usepackage{latexsym}
\usepackage{amsthm}

\usepackage[T1]{fontenc}

\usepackage[utf8]{inputenc}

\usepackage{microtype}

\usepackage{inconsolata}

\usepackage{graphicx}
\usepackage{enumitem}
\usepackage{amsmath}
\usepackage{algorithm}
\usepackage{algpseudocode}
\usepackage{subcaption} 

\usepackage{amsmath}
\usepackage{amssymb}
\usepackage{algorithm}
\usepackage{array}
\usepackage{multirow}
\usepackage{graphicx}
\title{\textbf{GraphRunner: A Multi-Stage Framework for Efficient and Accurate Graph-Based Retrieval}}

\author{Savini Kashmira \\
  University of Michigan \\
  \texttt{savinik@umich.edu} \\\And
  Jayanaka L. Dantanarayana \\
  University of Michigan \\
  \texttt{jayanaka@umich.edu} \\ \And
  Krisztián Flautner \\
  University of Michigan \\
  \texttt{manowar@umich.edu} \\ \AND
  Lingjia Tang \\
University of Michigan \\
  \texttt{lingjia@umich.edu} \\ \And
  Jason Mars \\
University of Michigan \\
  \texttt{profmars@umich.edu} \\
  }

\newcommand{\graphcot}[0]{G{\small RAPH}-C{\small O}T}
\newcommand{\runner}[0]{GraphRunner}
\newcommand{\grbench}[0]{GRB{\small ENCH}}

\begin{document}
\maketitle 
\begin{abstract}


Conventional Retrieval Augmented Generation (RAG) approaches are common in text-based applications. However, they struggle with structured, interconnected datasets like knowledge graphs, where understanding underlying relationships is crucial for accurate retrieval. A common direction in graph-based retrieval employs iterative, rule-based traversal guided by Large Language Models (LLMs). Such existing iterative methods typically combine reasoning with single hop traversal at each step, making them vulnerable to LLM reasoning errors and hallucinations that ultimately hinder the retrieval of relevant information.

To address these limitations, we propose \textbf{\runner{}}, a novel graph-based retrieval framework that operates in three distinct stages: planning, verification, and execution. This introduces high-level traversal actions that enable multi-hop exploration in a single step. It also generates a holistic traversal plan, which is verified against the graph structure and pre-defined traversal actions, reducing reasoning errors and detecting hallucinations before execution. 
Our evaluation using the \grbench{} dataset shows that \runner{} consistently outperforms existing approaches, achieving 10-50\% performance improvements over the strongest baseline while reducing inference cost by 3.0-12.9× and response generation time by 2.5-7.1×, making it significantly more robust and efficient for graph-based retrieval tasks.
\end{abstract}

\section{Introduction}

Large Language Models (LLMs) excel at understanding context and generating coherent text \cite{petroni2019language} but are prone to hallucinations, especially when lacking relevant information \cite{perkovic2024hallucinations}. To enhance the factuality and reliability of LLM outputs, Retrieval Augmented Generation (RAG) has emerged as state-of-the-art technique for external knowledge integration, substantially improving LLM performance~\cite{lewis2020retrieval, gao2024retrievalaugmentedgenerationlargelanguage}. It stores chunked documents in a vector database and retrieves the most relevant chunks based on query-chunk semantic similarity.



While RAG excels with unstructured text, it faces challenges with structured, interconnected datasets common in real-world scenarios. For instance, medical knowledge is inherently interlinked, diseases are associated with symptoms, drugs have side effects and treatments depend on anatomical factors. These relationships can be formally represented as knowledge graphs, where nodes encapsulate rich textual content and edges denote meaningful relationships. In such graph-based contexts, traditional RAG approaches fail to utilize structural connections effectively. These methods typically embed individual nodes and retrieve based only on vector similarity with the query, overlooking valuable relational information inherent in graphs which contain crucial context for answering complex queries.




To improve the retrieval from structured datasets, one of the prior approaches \cite{ye-etal-2024-language} leverages knowledge graphs by retrieving relevant nodes and fetching their single or multi-hop neighbors to form a subgraph, which is then linearized into a text sequence and passed to the LLM as context. While increasing the hop count can capture more relational context needed for complex queries, it can also lead to exponentially growing subgraphs that exceed the LLM's context window and include irrelevant information, consequently degrading performance \cite{liu-etal-2024-lost}. 





 To achieve better retrieval performance, rule-based graph traversal techniques are better suited than feeding entire subgraphs as context to LLMs. Existing frameworks \cite{jiang-etal-2023-structgpt, jin2024graphchainofthoughtaugmentinglarge} define traversal rules and leverage reasoning capabilities of LLMs to iteratively explore the graph. At each step, LLMs reason over the context, select the appropriate rule, and apply it.

However, these existing rule-based traversal techniques have several limitations. 
\begin{enumerate}[nosep]

    \item \textbf{Inefficient Graph Traversal Capabilities:} Since each iteration step only considers the immediate neighbors of the nodes retrieved in previous step, traversal is limited to single-hop exploration. Also, when identifying shared neighbors among multiple nodes, these frameworks retrieves all neighbors in one step and computes their intersection using LLMs in the next. As a result, they require multiple iteration steps for tasks that could be efficiently resolved using the graph structure effectively in fewer steps, leading to inefficiency. Relying on LLMs for such intermediate computations also increases the risk of LLM reasoning errors, reducing overall accuracy.

    \item \textbf{Step-by-Step Reasoning:} These frameworks combine traversal action planning and its execution into a single iterative step, requiring multiple reasoning steps to reach the final answer. Since each subsequent step is determined by LLMs based on previous reasoning steps and outputs, the amount of accumulated information grows with each step. This leads to higher token usage, reducing overall efficiency. Additionally, there is a risk of exceeding the model’s context window, hindering its ability to produce a final answer.

    \item \textbf{Undetected Hallucinations Before Execution:} LLMs hallucinate incorrect reasoning steps, such as suggesting traversal actions incompatible with pre-defined rules or referencing non-existent node/ edge types.  While we can sometimes identify them in the next reasoning step based on the previous execution feedback, the system often fails to detect them, leading to incorrect answers. Even when identified, corrections require longer traversals. Such hallucinations cannot be detected before execution in these frameworks.
    
    
    


    
\end{enumerate}

To address these limitations, we propose \textbf{\runner{}}, a novel three-stage graph-based retrieval framework that leverages LLMs for planning traversal steps based on pre-defined traversal actions, followed by verification and execution when the graph structure is defined. 
\runner{} introduces high-level traversal actions that perform complex operations such as multi hop traversal and shared neighbor identification in a single step, reducing LLM reasoning errors by eliminating LLM reliance, improving efficiency and accuracy. Additionally, unlike iterative approaches, \runner{} separates planning and execution into distinct stages, avoiding dependence on previous outputs to plan the next traversal action. Planning stage generates a holistic traversal plan in a single LLM inference, followed by verification of the plan's compatibility with the graph structure and defined traversal actions before execution. This detects hallucinations beforehand and avoids unnecessary long traversals, improving accuracy and efficiency.

We conduct experiments on \grbench{}~\cite{jin2024graphchainofthoughtaugmentinglarge} dataset to demonstrate the effectiveness of \runner{}, analyzing its performance against several baselines. Our evaluation results demonstrate that \runner{} significantly outperforms existing approaches in accuracy when answering diverse query types with varying levels of complexity, while also achieving lower inference cost and response generation speedup.

The main contributions of our paper are as follows.
\begin{enumerate}[nosep]
    \item A novel non-iterative framework, \runner{}, that separates planning from execution through verified traversal strategies. It introduces high-level traversal actions enabling complex operations in a single step, and holistic traversal planning with pre-execution verification against the graph structure and pre-defined actions.
    
    
    \item A comprehensive evaluation against several baselines, using the \grbench{} multi-domain benchmark dataset. Results show that \runner{} consistently outperforms all baselines across all domains, achieving 10-50\% improvements in GPT4Score over the strongest baseline, while simultaneously reducing inference costs by 3.0-12.9× and response generation time by 2.5-7.1×, demonstrating higher accuracy and efficiency for graph-based retrieval tasks.
\end{enumerate}

\begin{figure*}
    \centering
    \includegraphics[width=\linewidth]{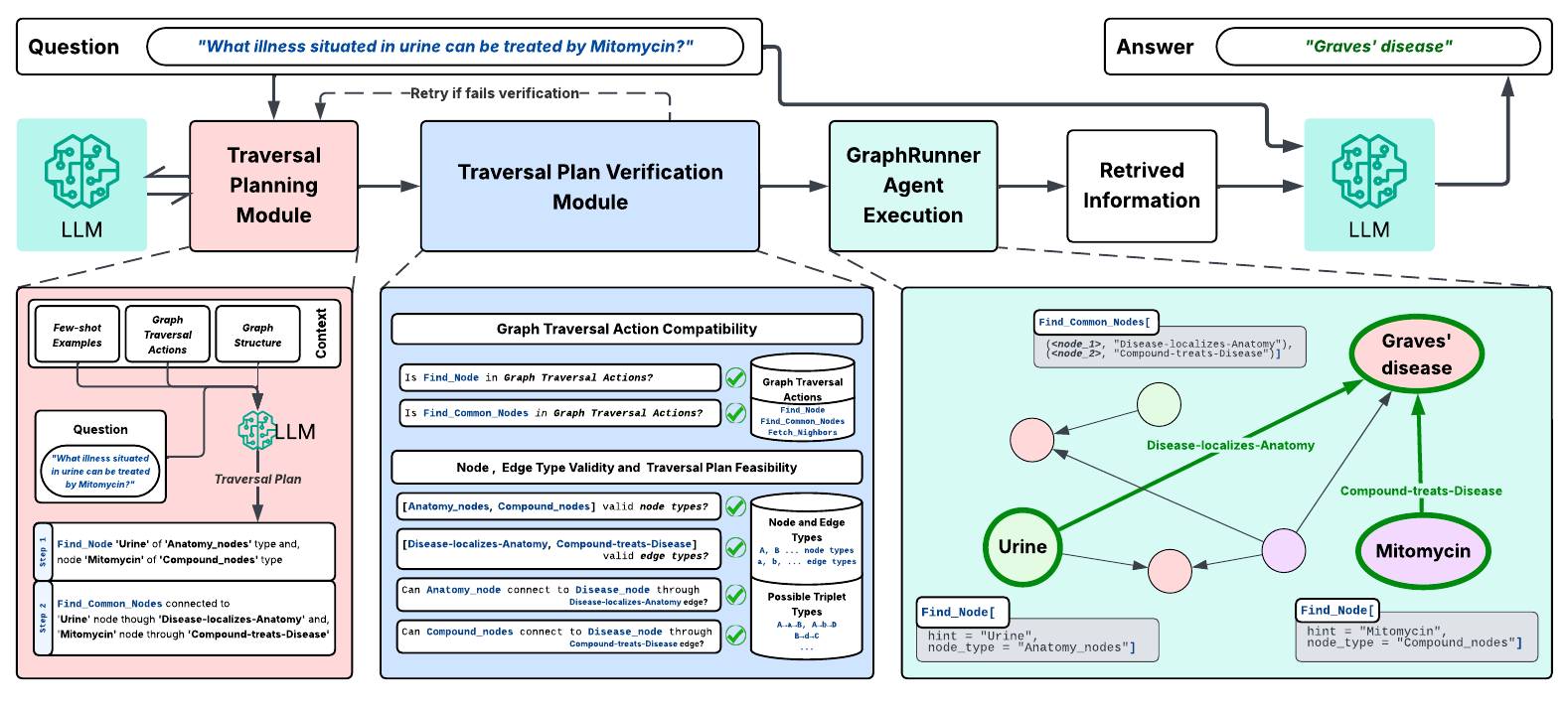}
    \caption{\runner{} Framework - Given a question, a holistic traversal plan is generated by the Traversal Planning module and verified against the graph and possible traversal actions in the Traversal Plan Verification Module. Once verified, the plan is executed to extract relevant information from the graph by the \runner{} Agent, which is then used as context to answer the question.}
    \label{fig:graphrunner}
    \vspace{-1em}
\end{figure*}

\section{Preliminaries}

This section formally introduces knowledge graphs and the mathematical representations underlying our implementations.

\subsection{Knowledge Graphs}

A knowledge graph is a structured representation of information that captures entities, their attributes, and the relationships between them. In our context, the knowledge graph is constructed with a known schema of node types and edge types.

\paragraph{Definition : Knowledge Graph.} Let $G = (V, E)$ be a knowledge graph where:
\begin{itemize} [nosep]
    \item $V$ - set of nodes and $E \subseteq V \times V$ - set of edges representing relationships between nodes.
    \item $T_v$ - set of node types (e.g., paper, author, venue in academic graphs) and $T_e$ - set of edge types (e.g., written-by, cites, published-in).
    \item Each node $v \in V$ has a type $\tau(v) \in T_v$ and a set of attributes $\text{attr}(v)$ (e.g., name, title, abstract, or other metadata).
    \item Each edge $e \in E$ has a type $\rho(e) \in T_e$.
\end{itemize}

\section{GraphRunner}

This section introduces GraphRunner, a novel graph-based retrieval framework shown in Figure \ref{fig:graphrunner}. This is a three-stage framework that begins by generating a holistic traversal plan, specifying which nodes and edges to traverse and how to navigate them using pre-defined traversal actions to answer the given query. In the second stage, the proposed traversal plan is verified against the actual graph structure and the capabilities of traversal agent before execution. In the final stage, the verified traversal plan is executed by the GraphRunner Agent to obtain the final answer.



The GraphRunner framework consists of three interconnected components: (1) GraphRunner Agent, (2) Traversal Planning Module, and (3) Traversal Plan Verification Module. 

\subsection{GraphRunner Agent} \label{subsec:GraphRunner_Agent}
The \textbf{GraphRunner Agent} serves as the core execution component that performs graph traversal operations as shown in Figure \ref{fig:graphrunner}. To systematically navigate knowledge graphs and retrieve relevant information, we define three graph traversal actions that the GraphRunner Agent can perform.

\subsubsection{Find\_Node Action}
\vspace{-1em}
\begin{figure}[H]
    \includegraphics[width=0.7\linewidth]{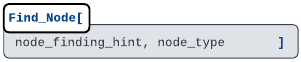}
\end{figure}
\vspace{-1em}
This action identifies specific nodes in the graph based on semantic similarity to the query, serving as the starting point for graph traversal. Once the nodes are found, the GraphRunner Agent begins traversal by visiting them.
We can formally define this action as:
\vspace{-0.6em}
\begin{equation*}
Find\_Node: H \times T_v \rightarrow \mathcal{P}(V_{\text{found}})
\end{equation*}

Where \( H \) is the set of node finding hints which extracted from the query and \( \mathcal{P}(V_{\text{found}}) \) represents the set of nodes matching the node finding hints. This action is implemented as:
\vspace{-0.6em}
\begin{equation*}
\resizebox{\columnwidth}{!}{
\ensuremath{
Find\_Node(h, t) = \{ v \in V_t \mid \text{\textit{similarity}}(h, \text{attr}(v)) \geq \theta \}
}}
\end{equation*}

Where \( V_t = \{v \in V : \tau(v) = t\} \) and \(\text{\textit{similarity}}\) is a semantic similarity function using vector embeddings, with \( \theta \) as a similarity threshold.





\subsubsection{Fetch\_Neighbors Action}
\vspace{-1em}
\begin{figure}[H]
    \includegraphics[width=0.7\linewidth]{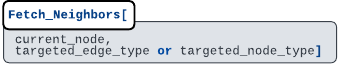}
    \vspace{-1em}
\end{figure}

This action enables the GraphRunner Agent to traverse the graph from the current node to its neighbors through specified relationships. The traversal can be either \textbf{single-hop} or \textbf{multi-hop}, depending on the parameters provided.

\begin{itemize}[nosep]
    \item For \textbf{single-hop traversal}, the agent uses \texttt{targeted\_edge\_type} as the argument. It fetches all immediate neighbors connected to the current node via edges of the given type.
    \item For \textbf{multi-hop traversal}, the agent uses the \texttt{targeted\_node\_type} as the argument. It repeatedly follows edges on the graph from the current node, until it reaches nodes matching the desired node type.
\end{itemize}

We can formally define this function as:
\begin{equation*}
\resizebox{\columnwidth}{!}{
\ensuremath{
Fetch\_Neighbors : V_{\text{current}} \times (T_e \cup T_v) \rightarrow \mathcal{P}(V_{\text{fetched}})
}}
\end{equation*}

The action is implemented as:
\begin{flushleft}
\begin{equation*}
\resizebox{0.6\columnwidth}{!}{
\ensuremath{
Fetch\_Neighbors(v, \text{\textit{param}}) = 
}}
\end{equation*}
\end{flushleft}
\begin{equation*}
\resizebox{0.95\columnwidth}{!}{
\ensuremath{
\begin{cases}
\{u \in V \mid (v,u) \in E, \rho((v,u)) = \text{\textit{param}}\}, & \text{if } \text{\textit{param}} \in T_e \\
\{u \in V \mid \exists \text{ path } p: v \to u, |p| \geq 1, \tau(u) = \text{\textit{param}}\}, & \text{if } \text{\textit{param}} \in T_v
\end{cases}
}}
\end{equation*}


Here, the implementation handles both cases depending on whether \textit{param} is an edge type (for single-hop traversal) or a node type (for multi-hop traversal). The path length $|p|$ indicates the number of hops from node $v$ to node $u$.





\subsubsection{Find\_Common\_Nodes Action}
\vspace{-1em}
\begin{figure}[H]
    \includegraphics[width=0.7\linewidth]{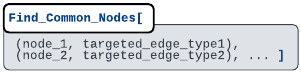}
    \vspace{-1em}
\end{figure}

This action can be used to find shared neighbors across different nodes. Figure~\ref{fig:graphrunner} illustrates an example of using this action.


We can formally define this function as:
\begin{equation*}
\resizebox{\columnwidth}{!}{
\ensuremath{
Find\_Common\_Nodes: (V \times T_e)^n \rightarrow \mathcal{P}(V_{\text{common}})
}}
\end{equation*}

The action is implemented as follows when the input tuples are \( (v_1, e_{t_1}), (v_2, e_{t_2}), \ldots, (v_n, e_{t_n}) \).
\begin{equation*}
\resizebox{\columnwidth}{!}{
\ensuremath{
Find\_Common\_Nodes((v_1, e_{t_1}), \ldots, (v_n, e_{t_n})) = \bigcap_{i=1}^{n} N_{e_{t_i}}(v_i)
}}
\end{equation*}
where, \begin{equation*}
\resizebox{\columnwidth}{!}{
\ensuremath{
N_{e_t}(v) = \{ u \in V : (v, u) \in E \text{ and } \rho((v, u)) = e_t \}
}}
\end{equation*}


\textit{By defining the high-level traversal actions: \texttt{Fetch\_Neighbors} and \texttt{Find\_Common\_Nodes}, \runner{} can perform multi-hop graph explorations in a single step.}


\subsection{Traversal Planning Module}

To determine how the GraphRunner Agent should navigate the graph using the defined graph traversal actions, we introduce the \textbf{Traversal Planning Module}. This generates a holistic traversal plan by leveraging the reasoning capabilities of LLMs.

This module constructs a prompt that includes: user query, graph structure description (specifications of node and edge types in the knowledge graph), detailed explanations of the defined three graph traversal actions and few-shot examples (demonstrations of how to construct traversal plans for various query types) as illustrated in Figure \ref{fig:graphrunner}.



Using this, the LLM performs reasoning to generate a comprehensive traversal plan. This plan strategically sequences the graph traversal steps required to retrieve relevant information from the graph. Each traversal step in this plan includes the necessary input parameters for the relevant graph traversal action, which the LLM determines based on the query and graph structure.

For instance, the starting action in any traversal plan is always the \texttt{Find\_Nodes} action, since we require initial nodes to begin traversal. This action requires two parameters: (1) \texttt{node\_finding\_hint} - a hint such as a node name or description about the node to find, that LLM derives from the query, (2) \texttt{node\_type} - the type of node to find. The LLM is capable of selecting them as it has context on query and graph structure. The LLM follows this same pattern for the remaining traversal actions, determining their input parameters, as illustrated in the Traversal Planning module in Figure~\ref{fig:graphrunner}.







\begin{table*}[ht]
\caption{Performance comparison of GraphRunner against all baseline methods across each domain in GRBENCH, evaluated using GPT4Score and R-L metrics.  The experiments were conducted using GPT-4.}
\label{tab:overall}
\centering
\resizebox{\textwidth}{!}{
\begin{tabular}{lcccccccccccccc}
& \multicolumn{2}{c}{\textit{Academic}} & & \multicolumn{2}{c}{\textit{E-Commerce}} & & \multicolumn{2}{c}{\textit{Healthcare}} & & \multicolumn{2}{c}{\textit{Legal}} & & \multicolumn{2}{c}{\textit{Literature}} \\

& GPT4Score & R-L && GPT4Score & R-L && GPT4Score & R-L && GPT4Score & R-L && GPT4Score & R-L \\
\cline{2-3} \cline{5-6} \cline{8-9} \cline{11-12} \cline{14-15}
&&&&&&&&&&&&&& \\
Text-based-RAG & 12.41 & 9.65 && 28.92 & 21.56 && 10.47 & 7.28 && 30.71 & 26.34 && 32.17 & 29.65 \\
Graph-based-RAG & 32.15 & 29.72 && 40.72 & 35.41 && 16.17 & 12.57 && 26.32 & 24.87 && 33.19 & 28.63 \\
Graph-CoT & 40.7 & 35.7 && 52.476 & 49.23 && 35.97 & 30.91 && 34.68 & 32.24 && 55.51 & 52.2 \\
\textbf{GraphRunner} & \textbf{61.25} & \textbf{59.05} && \textbf{58.00} & \textbf{55.36} && \textbf{52.66} & \textbf{53.46} && \textbf{44.95} & \textbf{44.38} && \textbf{61.52} & \textbf{60.21} \\
\hline
\end{tabular}
}
\vspace{-1em}
\end{table*}

\subsection{Traversal Plan Verification Module}

Since LLMs are prone to hallucination, the generated graph traversal plan may reference non-existent graph elements or be otherwise non-executable. To mitigate this, we introduce the \textbf{Traversal Plan Verification Module}, which ensures that the generated traversal plan is valid and executable before passing to the GraphRunner Agent for execution.

The Traversal Plan Verification Module performs two key validation steps:
\begin{enumerate}[nosep]
    \item \textbf{Traversal Action Compatibility}: This step verifies that all traversal actions in the plan are among the pre-defined set of traversal actions supported by the GraphRunner Agent.
    \item \textbf{Graph Structure  Compatibility and Feasibility}: In this step, we validate the sequence of traversal actions in the plan, step by step. For each action, we check whether the specified input parameters such as node/ edge types exist in the graph structure. In particular, if an edge type is specified, we ensure that it exists as a valid connection from the given node to traverse, as illustrated in the Traversal Plan Verification Module in Figure~\ref{fig:graphrunner}.
\end{enumerate}

If any validation step fails, the system routes the plan back to the Traversal Planning Module to regenerate a revised version that is structurally valid. The verified traversal plan is forwarded to the GraphRunner Agent for execution only after all validation checks have passed.

\subsection{Final Answer Generation}

After receiving the verified traversal plan, the GraphRunner Agent executes each traversal action one-at-a-time, following the specified order in the plan. The destination node resulting from one traversal action serves as the input for the next. Once the final traversal action is completed, the agent takes its output and passes it to the LLM along with the original query to generate the final answer.

\section{Experiments}

\subsection{Experimental Setup}
\paragraph{Dataset.} We evaluate \runner{} and baseline approaches using the \grbench{} dataset \cite{jin2024graphchainofthoughtaugmentinglarge}, which contains graphs from 5 domains and a total of 1,740 questions categorized into three difficulty levels: easy, medium, and hard, in English Language. However, in the healthcare domain, only easy and medium questions are available. This dataset is Apache-2.0 licensed on HuggingFace.
\vspace{-0.5em}

\paragraph{Baselines.} We evaluate our \runner{} approach against several baseline approaches. 
\begin{itemize}[nosep]
    \item \textbf{Text-based RAG:} This retrieves relevant individual nodes from the dataset using a RAG framework. The retrieved nodes are directly passed to the LLM \cite{lewis2021retrievalaugmentedgenerationknowledgeintensivenlp}.
    
    \item \textbf{Graph-based RAG:} Building on text-based RAG, this first retrieves relevant nodes and then expands the context by including their multi-hop neighbors to construct a subgraph and linearize into a text sequence, which is then processed by the LLM. \cite{ye-etal-2024-language}. Here, we consider 2-hop neighbors.
    
    \item \textbf{\graphcot{}:} This augments LLMs with graph-based reasoning by allowing the model to iteratively reason and traverse over the graph~\cite{jin2024graphchainofthoughtaugmentinglarge}. At each step, it explores the neighbors of the current node, performing only 1-hop traversal to progressively gather relevant information. (Source code is Apache-2.0 licensed on GitHub.)
\end{itemize}

\paragraph{Evaluation Metrics.} To evaluate performance, we use two main metrics. GPT4Score \cite{fu2023gptscore} measures the percentage of "correct answer" predicted by GPT-4 that is relative to the ground truth answer. Rogue-L (R-L) \cite{lin-2004-rouge} evaluates the percentage overlap between the longest common subsequence of words between the generated response and the ground truth answer.

To evaluate efficiency, we compute the average inference cost per question using input and output tokens, as defined by OpenAI for GPT-4 model~\cite{openai_pricing_other_models}.
\begin{equation*}
\resizebox{0.9\columnwidth}{!}{
\ensuremath{
    \textit{Cost} = \$30 \times \left( \frac{\textit{Input Tokens}}{1M} \right) + 
                  \$60 \times \left( \frac{\textit{Output Tokens}}{1M} \right)
}}
\end{equation*}
We also report end-to-end response time as another metric for efficiency.


\begin{figure*}




    \centering
    \includegraphics[width=\linewidth]{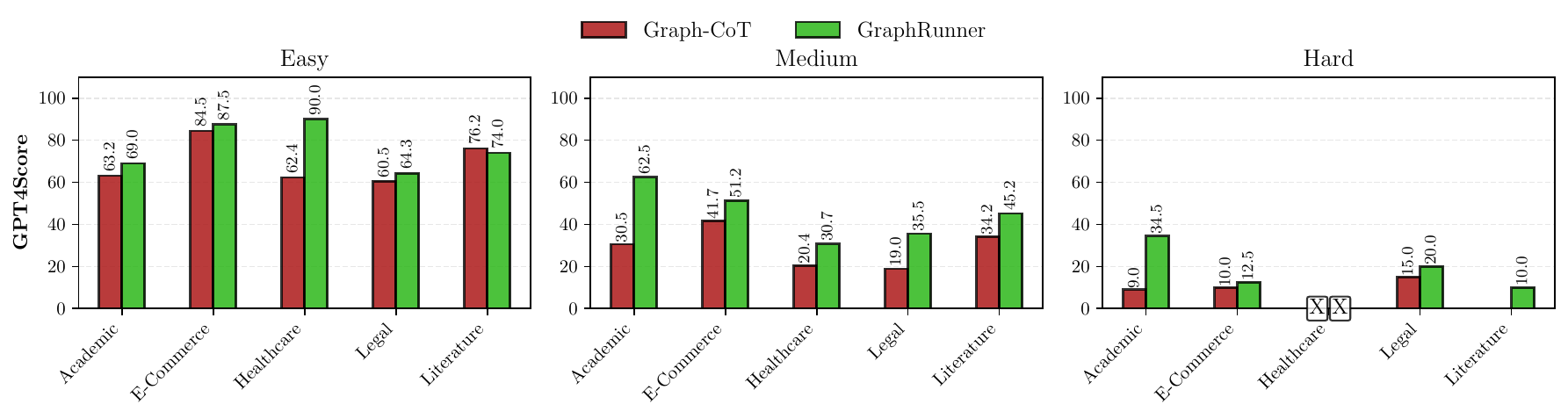}
    \caption{Performance improvements of \runner{} over \graphcot{} across all domains and difficulty levels in the \grbench{} dataset, evaluated using GPT4Score. The experiments were conducted using GPT-4. Note that only easy and medium difficulty questions are available for the healthcare domain.}
    \label{fig:gpt4score}
    \vspace{-1em}
    
\end{figure*}

\subsection{Retrieval Performance Evaluation}

We compare \runner{} against all baseline methods using \grbench{} datasets for all the domains. As summarized in Table~\ref{tab:overall}, \runner{} consistently outperforms all baselines methods evaluated using GPT-4.

Text-based RAG performs poorly as it retrieves only individual nodes without capturing essential relationships between nodes. As evident from Table \ref{tab:overall}, Graph-based RAG improves over text-based RAG by constructing a 2-hop subgraph. However, connecting all neighbors regardless of edge type introduces noisy data and can exceed the context window due to rapid subgraph growth, causing it to under-perform compared to other baselines. \graphcot{} improves over RAG methods by using rule based traversal, guided by LLM reasoning, to decide which rule to apply at each iteration.

\begin{figure*}[ht]
    \centering
    \begin{subfigure}[b]{0.93\columnwidth}
        \centering
        \includegraphics[width=\linewidth]{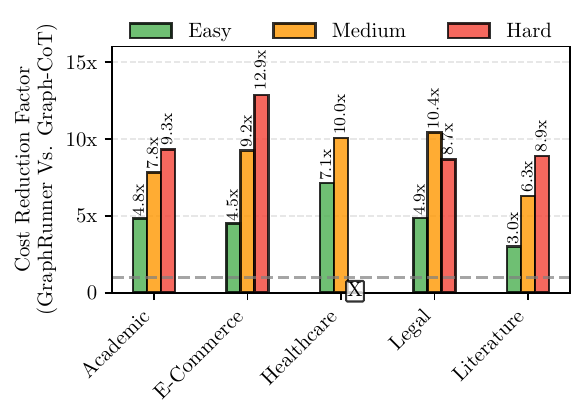}
        \caption{}
        \label{fig:cost}
    \end{subfigure}
    \begin{subfigure}[b]{0.93\columnwidth}
        \centering
        \includegraphics[width=\linewidth]{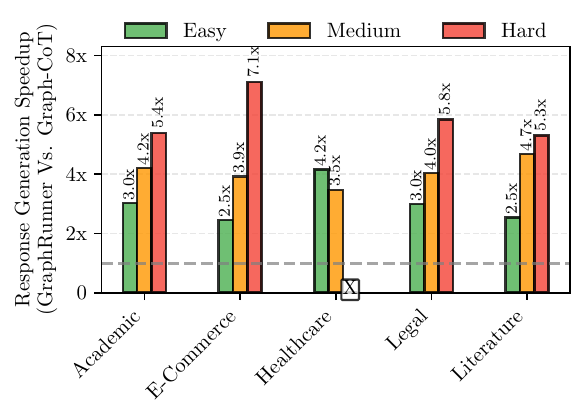}
        \caption{}
        \label{fig:speedup_latency}
    \end{subfigure}
    \caption{Efficiency improvements of \runner{} over \graphcot{} across all domains and difficulty levels in the \grbench{} dataset. (a) Cost Reduction Factor: the ratio of \graphcot{}’s LLM inference cost to that of \runner{}, indicating significant cost savings. (b) Response Generation Speedup: the ratio of \graphcot{}’s end-to-end response generation time to that of \runner{}, showing notable improvements in response time. Note that only easy and medium difficulty questions are available for the healthcare domain.}
    \label{fig:efficiency}
    \vspace{-1em}
\end{figure*}

However, \runner{} consistently outperforms all baselines by addressing their limitations. While \graphcot{} requires LLMs to reason through multiple steps to fetch multi-hop neighbors and find shared neighbors, \runner{} introduces high-level traversal actions that accomplish these in a single step, reducing reasoning burden for LLMs, consequently minimizing likelihood of reasoning errors. Furthermore, \runner{} generates and verifies a complete traversal plan against the graph structure and possible traversal actions, detecting hallucinations before execution and enhancing performance. These enhancements lead to performance gains of 10-50\% over \graphcot{}, the strongest baseline according to the Table \ref{tab:overall}.

\paragraph{Performance Across Different Question Difficulty Levels in \grbench{}.}
We conducted an experiment to check how \runner{} is performing for questions with various difficulty levels: easy, medium, and hard, compared to \graphcot{}. As shown in Figure \ref{fig:gpt4score}, \runner{} outperforms \graphcot{} across all difficulty levels: marginal improvements for easy questions, significant improvements for medium questions, and moderate improvements for hard questions achieving higher GPT4Scores. This trend can be explained using how different question types interact with graph structures and how traversal actions are leveraged.
 
Easy questions typically require only single-node lookups or 1-hop traversal. Since they do not involve complex reasoning and traversal, the performance improvement mainly comes from \runner{}'s ability to reduce errors due to hallucinations through validated traversal planning before execution. We see a significant performance improvement of around 44\% in healthcare domain, which has the most complex graph with 11 node types and 24 edge types. This structural complexity increases the chance of LLM reasoning errors and hallucinations, which \runner{} mitigates through traversal plan validation. 
This highlights the effectiveness of \runner{} in answering for easy questions even over highly complex graph structures due to the verification stage.

Medium questions require multi-hop traversals and finding shared neighbors across different nodes. \runner{}'s high-level traversal actions allow it to execute these complex traversal operations within few steps than \graphcot{}. In addition to validation of traversal plan before execution, this leads to significant performance improvements ranging from 20-90\% across all domains highlighting the importance of introducing these high-level traversal actions.



Hard questions in the dataset are, by definition, "\textit{questions that cannot be directly answered by looking up the graph, but the graph can be useful by providing informative context}"~\cite{jin2024graphchainofthoughtaugmentinglarge}. For example, paper recommendation questions for a given reader based on a previously read paper in the academic dataset, lack a specific traversal path to fetch a recommended paper for a given reader, as connections may exist through multiple paths (citations, venues, authors) or through conceptual similarities not captured by the graph. When a system produces correct answers, relevant nodes often appear coincidentally along the traversal path. Nevertheless, \runner{} shows a modest improvement over \graphcot{} for hard questions by leveraging high-level traversal actions that enable more complex reasoning and increase the likelihood of finding relevant nodes, especially in academic and literature graphs as shown in Figure \ref{fig:gpt4score}.

\subsection{Retrieval Efficiency Evaluation}

\begin{figure*}
    \centering
    \includegraphics[width=0.85\linewidth]{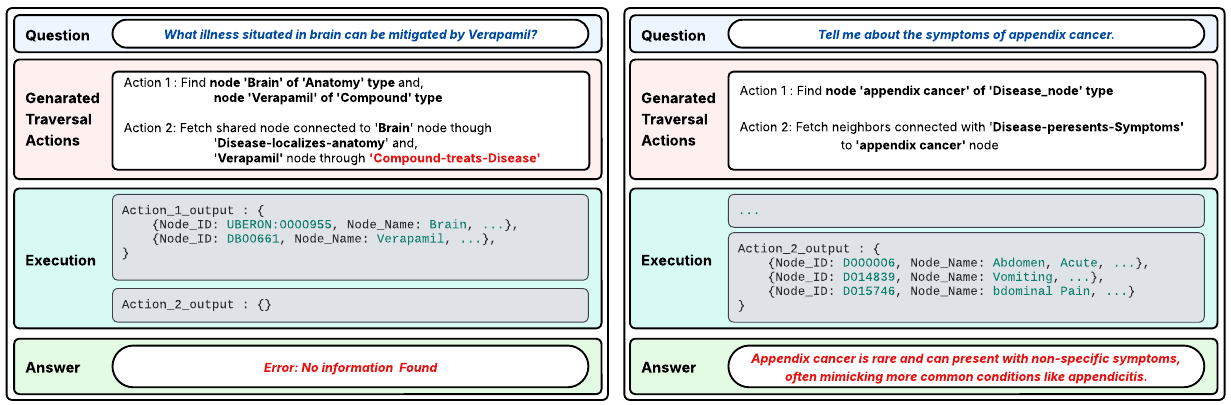}
    \caption{Examples for failure cases in \runner{}: the first example on the left demonstrates a reasoning error and the second example on the right highlights a hallucination error.}
    \label{fig:error_example}
    \vspace{-0.8em}
\end{figure*}

While \runner{} demonstrates significant improvements in retrieval performance, assessing its efficiency is equally important. We conduct a comprehensive efficiency analysis comparing \runner{} with \graphcot{}, including both inference cost measurement and response generation time evaluation.

Figure~\ref{fig:cost} illustrates the inference cost savings while Figure~\ref{fig:speedup_latency} shows the response generation speedup achieved by \runner{} compared to \graphcot{} across all domains and difficulty levels in the \grbench{} dataset. Cost savings are reported using the Cost Reduction Factor, defined as the ratio of \graphcot{}'s LLM token usage cost to \runner{}'s. Speedup is measured as the ratio of \graphcot{}'s end-to-end response generation time to \runner{}'s.

\runner{} consistently demonstrates significant reductions in inference cost, with reduction factors ranging from \textbf{3.0× - 12.9×} across all domains. Simultaneously, it produces faster responses across all domains and difficulty levels, with speedups ranging from \textbf{2.5× - 7.1×}. These findings demonstrate that \runner{} not only reduces inference costs but also significantly accelerates response times without sacrificing performance. These efficiency improvements can be attributed to two key innovations in \runner{}: separation of planning from execution and introduction of high-level traversal actions.


\runner{} generates a complete traversal plan in a single inference, followed by verification before execution. If verification fails and triggers a retry, only the failed step is added as feedback to the next inference to correct the plan, minimizing token usage. After executing the plan, the retrieved information from the final traversal action is fed back to the LLM with the original question to produce the final answer in one more inference. In contrast, \graphcot{} performs multiple inferences for iterative reasoning, causing token usage to grow with each accumulated step and output. Thus, \runner{} uses only a fraction of the tokens compared to \graphcot{}, significantly reducing cost and response generation time.




By implementing high-level traversal actions, \runner{} completes complex operations in fewer steps, reducing the total tokens needed to generate the traversal plan. In contrast, \graphcot{} requires more reasoning steps to achieve similar results, leading to a higher token count and increased inference cost. \runner{} not only reduces total token usage but is especially effective at minimizing output tokens by shortening the traversal plan. Since inference latency in auto-regressive transformer models such as GPT-4 increases approximately quadratically with output token length~\cite{chen2024sequoia}, reducing output tokens in \runner{} significantly accelerates response generation time.



\subsection{Case Study for Error Analysis.}

We conducted a case study to investigate the failure cases of \runner{} and how these errors are reduced compared with \graphcot{}.


Figure~\ref{fig:error_example} shows two cases where \runner{} fails. In the left example, the LLM mistakenly selects the edge type \textit{Compound palliates Disease} instead of \textit{Compound treats Disease}. The Traversal Verification Module validates the plan based only on the structural correctness, ignoring the question's semantic intent. As a result, the traversal uses incorrect relationships, causing it to break and fail to produce a final answer.


In the example on the right, the verified traversal plan executes successfully, but the LLM fabricates a final answer without using the retrieved context, a hallucination error. While verification helps reduce such errors during plan generation, they can still occur during answer generation in \runner{}.

We manually analyzed such failure cases in \runner{} and classified them into three types:
\begin{enumerate}[nosep]
\item \textbf{Hallucination Errors}: Occur when the LLM fabricates incorrect reasoning steps or final answers.
\item \textbf{Reasoning Errors}: Occur when the traversal complies with the graph structure and pre-defined traversal rules, but the semantic intent of the traversal is incorrect, as discussed in the left example of Figure \ref{fig:error_example}.
\item \textbf{Context Window Exceeding Errors}: Occur when an LLM inference is invoked with an input prompt whose token count exceeds the context window size of the LLM in use.
\end{enumerate}
\begin{table}[h]
    \caption{Probability of each error type occurring in responses generated by \runner{} and \graphcot{} across all 1,740 questions in \grbench{} dataset.}
    \label{tab:llm_error_distribution}
    \centering
    \resizebox{0.98\linewidth}{!}{
    \begin{tabular}{lcccc}
        \multicolumn{1}{c}{\textit{}} & & \multicolumn{3}{c}{\textbf{Error Occurrence Probability (\%)}} \\
        \multicolumn{1}{c}{\textit{\textbf{Error Type}}} & &\textit{\graphcot{}} & & \textit{\runner{}} \\
        \cline{1-1} \cline{3-3} \cline{5-5}
        &&&& \\
        Hallucination errors      & & 14.12 & & 7.75 \\
        Reasoning errors      & & 33.97 & & 29.85 \\
        Context window exceeding errors & & 5.21 & & 1.16 \\
        \hline
    \end{tabular}
    }
\vspace{-0.3em}
\end{table}
After categorizing all failure cases of \runner{} and \graphcot{} into these three types, we computed the probability of each error type occurring across the responses generated for all 1,740 questions in the \grbench{} dataset. The results are summarized in Table~\ref{tab:llm_error_distribution}. 

Our results show that nearly 50\% of hallucination errors in \graphcot{} are reduced by separating traversal planning and verification from execution in \runner{}. By verifying that the generated plan aligns with the graph structure and pre-defined traversal actions, we can detect hallucinated actions or invalid node or edge types before execution, reducing hallucination errors. However, as shown in the left example of Figure \ref{fig:error_example}, \runner{} can still hallucinate during final answer generation, although to a lesser extent. This highlights the importance of the Traversal Path Verification Module in minimizing hallucinations.

In terms of reasoning errors, we observe a 12\% reduction in error probability with \runner{} compared to Graph CoT. As defined earlier, these errors mainly stem from limitations in the reasoning capabilities of LLMs. However, by introducing high-level traversal actions, \runner{} reduces reliance on the LLM's reasoning capability, thereby lowering the likelihood of such errors.

There is a significant reduction in context window exceeding error occurrence probability with \runner{} relative to \graphcot{}, approximately around 80\%. \graphcot{} experiences these errors mainly due to the accumulation of previous reasoning steps and their outputs given as input to the current step. In contrast, \runner{} generates a holistic traversal plan ideally with one inference, requiring fewer steps to accomplish the same traversal due to high-level traversal actions. Context window exceeding errors can still occur when performing the final inference based on the output of the last traversal action in \runner{}, but this only occurs in about 1\% of responses.

Through this case study, we highlight how \runner{} reduces error occurrence probability across different types of errors and to what degree, through the introduction of high-level traversal actions and verified traversal.

\section{Related Work}
\vspace{-0.5 em}

Recent work shows growing interest in LLM-powered graph tasks \cite{10.1145/3589335.3651476,10.1145/3655103.3655110, zhang2023graphtoolformerempowerllmsgraph, wu2024can, liudual}, with integration advancing in several directions. One direction converts graph data into textual input for LLMs, as in InstructGLM \cite{ye-etal-2024-language} and NLGraph \cite{wang2023can}. Another combines Graph Neural Networks (GNNs) with LLMs, as in GraphGPT \cite{10.1145/3626772.3657775} and GraphLLM \cite{chai2023graphllm}. These approaches provide full graph context and often require fine-tuning. In contrast, our work retrieves relevant context and uses pre-trained LLMs without additional training.

Recent works have extended RAG techniques \cite{lewis2020retrieval, jiang-etal-2023-active, chen2024benchmarking} to better handle graph-based datasets \cite{10.1145/3626772.3661370, he2024g, jiang2024ragraph}. GraphRAG \cite{edge2024local} constructs a knowledge graph from text, summarizes node communities, and aggregates partial responses to produce a final answer. However, its focus is more on query-focused summarization than general retrieval. Some methods use rule-based graph traversal guided by LLM reasoning, with iterative frameworks like StructGPT \cite{jiang-etal-2023-structgpt} and Graph-CoT \cite{jin2024graphchainofthoughtaugmentinglarge} combining reasoning and execution at each step. In contrast, we propose a rule-based approach that separates LLM-driven traversal planning from execution.



\section{Conclusion}
\vspace{-0.5 em}

We present \runner{}, a three-stage framework: traversal planning, plan verification, and execution for graph-based retrieval. This enables multi-hop graph exploration via high-level traversal actions in a single step. By separating planning and execution, token usage can be reduced. The verification stage detects hallucinations before execution. Experiments on the \grbench{} dataset shows that \runner{} consistently outperforms existing methods in both performance and efficiency. 

\section*{Limitations}

Our implementation of \runner{} relies on OpenAI's proprietary GPT-4 API, which has shown strong performance on reasoning tasks. However, a key limitation of our work is the lack of evaluation using open source LLMs with fine-tuning capabilities, an avenue that can be explored in future research. Furthermore, our evaluation was constrained by the \grbench{} dataset, which, despite containing knowledge graphs with question-answer pairs, offers limited question diversity. A more robust evaluation would be possible if newer datasets emerge with more comprehensive question-answer sets and corresponding knowledge graphs.



\bibliography{references}

\begin{thebibliography}{27}
\providecommand{\natexlab}[1]{#1}

\bibitem[{Chai et~al.(2023)Chai, Zhang, Wu, Han, Hu, Huang, and Yang}]{chai2023graphllm}
Ziwei Chai, Tianjie Zhang, Liang Wu, Kaiqiao Han, Xiaohai Hu, Xuanwen Huang, and Yang Yang. 2023.
\newblock Graphllm: Boosting graph reasoning ability of large language model.
\newblock \emph{arXiv preprint arXiv:2310.05845}.

\bibitem[{Chen et~al.(2024{\natexlab{a}})Chen, Lin, Han, and Sun}]{chen2024benchmarking}
Jiawei Chen, Hongyu Lin, Xianpei Han, and Le~Sun. 2024{\natexlab{a}}.
\newblock Benchmarking large language models in retrieval-augmented generation.
\newblock In \emph{Proceedings of the AAAI Conference on Artificial Intelligence}, volume~38, pages 17754--17762.

\bibitem[{Chen et~al.(2024{\natexlab{b}})Chen, Mao, Li, Jin, Wen, Wei, Wang, Yin, Fan, Liu, and Tang}]{10.1145/3655103.3655110}
Zhikai Chen, Haitao Mao, Hang Li, Wei Jin, Hongzhi Wen, Xiaochi Wei, Shuaiqiang Wang, Dawei Yin, Wenqi Fan, Hui Liu, and Jiliang Tang. 2024{\natexlab{b}}.
\newblock \href {https://doi.org/10.1145/3655103.3655110} {Exploring the potential of large language models (llms)in learning on graphs}.
\newblock \emph{SIGKDD Explor. Newsl.}, 25(2):42–61.

\bibitem[{Chen et~al.(2024{\natexlab{c}})Chen, May, Svirschevski, Huang, Ryabinin, Jia, and Chen}]{chen2024sequoia}
Zhuoming Chen, Avner May, Ruslan Svirschevski, Yu-Hsun Huang, Max Ryabinin, Zhihao Jia, and Beidi Chen. 2024{\natexlab{c}}.
\newblock Sequoia: Scalable and robust speculative decoding.
\newblock \emph{Advances in Neural Information Processing Systems}, 37:129531--129563.

\bibitem[{Edge et~al.(2024)Edge, Trinh, Cheng, Bradley, Chao, Mody, Truitt, Metropolitansky, Ness, and Larson}]{edge2024local}
Darren Edge, Ha~Trinh, Newman Cheng, Joshua Bradley, Alex Chao, Apurva Mody, Steven Truitt, Dasha Metropolitansky, Robert~Osazuwa Ness, and Jonathan Larson. 2024.
\newblock From local to global: A graph rag approach to query-focused summarization.
\newblock \emph{arXiv preprint arXiv:2404.16130}.

\bibitem[{Fu et~al.(2023)Fu, Ng, Jiang, and Liu}]{fu2023gptscore}
Jinlan Fu, See-Kiong Ng, Zhengbao Jiang, and Pengfei Liu. 2023.
\newblock Gptscore: Evaluate as you desire.
\newblock \emph{arXiv preprint arXiv:2302.04166}.

\bibitem[{Gao et~al.(2024)Gao, Xiong, Gao, Jia, Pan, Bi, Dai, Sun, Wang, and Wang}]{gao2024retrievalaugmentedgenerationlargelanguage}
Yunfan Gao, Yun Xiong, Xinyu Gao, Kangxiang Jia, Jinliu Pan, Yuxi Bi, Yi~Dai, Jiawei Sun, Meng Wang, and Haofen Wang. 2024.
\newblock \href {https://arxiv.org/abs/2312.10997} {Retrieval-augmented generation for large language models: A survey}.
\newblock \emph{Preprint}, arXiv:2312.10997.

\bibitem[{He et~al.(2024)He, Tian, Sun, Chawla, Laurent, LeCun, Bresson, and Hooi}]{he2024g}
Xiaoxin He, Yijun Tian, Yifei Sun, Nitesh Chawla, Thomas Laurent, Yann LeCun, Xavier Bresson, and Bryan Hooi. 2024.
\newblock G-retriever: Retrieval-augmented generation for textual graph understanding and question answering.
\newblock \emph{Advances in Neural Information Processing Systems}, 37:132876--132907.

\bibitem[{Jiang et~al.(2023{\natexlab{a}})Jiang, Zhou, Dong, Ye, Zhao, and Wen}]{jiang-etal-2023-structgpt}
Jinhao Jiang, Kun Zhou, Zican Dong, Keming Ye, Xin Zhao, and Ji-Rong Wen. 2023{\natexlab{a}}.
\newblock \href {https://doi.org/10.18653/v1/2023.emnlp-main.574} {{S}truct{GPT}: A general framework for large language model to reason over structured data}.
\newblock In \emph{Proceedings of the 2023 Conference on Empirical Methods in Natural Language Processing}, pages 9237--9251, Singapore. Association for Computational Linguistics.

\bibitem[{Jiang et~al.(2024)Jiang, Qiu, Xu, Zhu, Zhang, Fang, Xu, Zhao, and Wang}]{jiang2024ragraph}
Xinke Jiang, Rihong Qiu, Yongxin Xu, Yichen Zhu, Ruizhe Zhang, Yuchen Fang, Chu Xu, Junfeng Zhao, and Yasha Wang. 2024.
\newblock Ragraph: A general retrieval-augmented graph learning framework.
\newblock \emph{Advances in Neural Information Processing Systems}, 37:29948--29985.

\bibitem[{Jiang et~al.(2023{\natexlab{b}})Jiang, Xu, Gao, Sun, Liu, Dwivedi-Yu, Yang, Callan, and Neubig}]{jiang-etal-2023-active}
Zhengbao Jiang, Frank Xu, Luyu Gao, Zhiqing Sun, Qian Liu, Jane Dwivedi-Yu, Yiming Yang, Jamie Callan, and Graham Neubig. 2023{\natexlab{b}}.
\newblock \href {https://doi.org/10.18653/v1/2023.emnlp-main.495} {Active retrieval augmented generation}.
\newblock In \emph{Proceedings of the 2023 Conference on Empirical Methods in Natural Language Processing}, pages 7969--7992, Singapore. Association for Computational Linguistics.

\bibitem[{Jin et~al.(2024)Jin, Xie, Zhang, Roy, Zhang, Li, Li, Tang, Wang, Meng, and Han}]{jin2024graphchainofthoughtaugmentinglarge}
Bowen Jin, Chulin Xie, Jiawei Zhang, Kashob~Kumar Roy, Yu~Zhang, Zheng Li, Ruirui Li, Xianfeng Tang, Suhang Wang, Yu~Meng, and Jiawei Han. 2024.
\newblock \href {https://arxiv.org/abs/2404.07103} {Graph chain-of-thought: Augmenting large language models by reasoning on graphs}.
\newblock \emph{Preprint}, arXiv:2404.07103.

\bibitem[{Lewis et~al.(2020)Lewis, Perez, Piktus, Petroni, Karpukhin, Goyal, K{\"u}ttler, Lewis, Yih, Rockt{\"a}schel et~al.}]{lewis2020retrieval}
Patrick Lewis, Ethan Perez, Aleksandra Piktus, Fabio Petroni, Vladimir Karpukhin, Naman Goyal, Heinrich K{\"u}ttler, Mike Lewis, Wen-tau Yih, Tim Rockt{\"a}schel, and 1 others. 2020.
\newblock Retrieval-augmented generation for knowledge-intensive nlp tasks.
\newblock \emph{Advances in neural information processing systems}, 33:9459--9474.

\bibitem[{Lewis et~al.(2021)Lewis, Perez, Piktus, Petroni, Karpukhin, Goyal, Küttler, Lewis, tau Yih, Rocktäschel, Riedel, and Kiela}]{lewis2021retrievalaugmentedgenerationknowledgeintensivenlp}
Patrick Lewis, Ethan Perez, Aleksandra Piktus, Fabio Petroni, Vladimir Karpukhin, Naman Goyal, Heinrich Küttler, Mike Lewis, Wen tau Yih, Tim Rocktäschel, Sebastian Riedel, and Douwe Kiela. 2021.
\newblock \href {https://arxiv.org/abs/2005.11401} {Retrieval-augmented generation for knowledge-intensive nlp tasks}.
\newblock \emph{Preprint}, arXiv:2005.11401.

\bibitem[{Lin(2004)}]{lin-2004-rouge}
Chin-Yew Lin. 2004.
\newblock \href {https://aclanthology.org/W04-1013/} {{ROUGE}: A package for automatic evaluation of summaries}.
\newblock In \emph{Text Summarization Branches Out}, pages 74--81, Barcelona, Spain. Association for Computational Linguistics.

\bibitem[{Liu et~al.()Liu, Zhang, Li, and Yao}]{liudual}
Guangyi Liu, Yongqi Zhang, Yong Li, and Quanming Yao.
\newblock Dual reasoning: A gnn-llm collaborative framework for knowledge graph question answering.
\newblock In \emph{The Second Conference on Parsimony and Learning (Proceedings Track)}.

\bibitem[{Liu et~al.(2024{\natexlab{a}})Liu, Lin, Hewitt, Paranjape, Bevilacqua, Petroni, and Liang}]{liu-etal-2024-lost}
Nelson~F. Liu, Kevin Lin, John Hewitt, Ashwin Paranjape, Michele Bevilacqua, Fabio Petroni, and Percy Liang. 2024{\natexlab{a}}.
\newblock \href {https://doi.org/10.1162/tacl_a_00638} {Lost in the middle: How language models use long contexts}.
\newblock \emph{Transactions of the Association for Computational Linguistics}, 12:157--173.

\bibitem[{Liu et~al.(2024{\natexlab{b}})Liu, He, Tian, and Chawla}]{10.1145/3589335.3651476}
Zheyuan Liu, Xiaoxin He, Yijun Tian, and Nitesh~V. Chawla. 2024{\natexlab{b}}.
\newblock \href {https://doi.org/10.1145/3589335.3651476} {Can we soft prompt llms for graph learning tasks?}
\newblock In \emph{Companion Proceedings of the ACM Web Conference 2024}, WWW '24, page 481–484, New York, NY, USA. Association for Computing Machinery.

\bibitem[{{OpenAI}(2025)}]{openai_pricing_other_models}
{OpenAI}. 2025.
\newblock Openai api pricing: Other models.
\newblock \url{https://platform.openai.com/docs/pricing#other-models}.
\newblock Accessed: 2025-05-19.

\bibitem[{Perkovi{\'c} et~al.(2024)Perkovi{\'c}, Drobnjak, and Boti{\v{c}}ki}]{perkovic2024hallucinations}
Gabrijela Perkovi{\'c}, Antun Drobnjak, and Ivica Boti{\v{c}}ki. 2024.
\newblock Hallucinations in llms: Understanding and addressing challenges.
\newblock In \emph{2024 47th MIPRO ICT and Electronics Convention (MIPRO)}, pages 2084--2088. IEEE.

\bibitem[{Petroni et~al.(2019)Petroni, Rockt{\"a}schel, Lewis, Bakhtin, Wu, Miller, and Riedel}]{petroni2019language}
Fabio Petroni, Tim Rockt{\"a}schel, Patrick Lewis, Anton Bakhtin, Yuxiang Wu, Alexander~H Miller, and Sebastian Riedel. 2019.
\newblock Language models as knowledge bases?
\newblock \emph{arXiv preprint arXiv:1909.01066}.

\bibitem[{Tang et~al.(2024)Tang, Yang, Wei, Shi, Su, Cheng, Yin, and Huang}]{10.1145/3626772.3657775}
Jiabin Tang, Yuhao Yang, Wei Wei, Lei Shi, Lixin Su, Suqi Cheng, Dawei Yin, and Chao Huang. 2024.
\newblock \href {https://doi.org/10.1145/3626772.3657775} {Graphgpt: Graph instruction tuning for large language models}.
\newblock In \emph{Proceedings of the 47th International ACM SIGIR Conference on Research and Development in Information Retrieval}, SIGIR '24, page 491–500, New York, NY, USA. Association for Computing Machinery.

\bibitem[{Wang et~al.(2023)Wang, Feng, He, Tan, Han, and Tsvetkov}]{wang2023can}
Heng Wang, Shangbin Feng, Tianxing He, Zhaoxuan Tan, Xiaochuang Han, and Yulia Tsvetkov. 2023.
\newblock Can language models solve graph problems in natural language?
\newblock \emph{Advances in Neural Information Processing Systems}, 36:30840--30861.

\bibitem[{Wu et~al.(2024)Wu, Shen, Shan, Song, Wang, Zhang, Feng, Cheng, Chen, Xiong et~al.}]{wu2024can}
Xixi Wu, Yifei Shen, Caihua Shan, Kaitao Song, Siwei Wang, Bohang Zhang, Jiarui Feng, Hong Cheng, Wei Chen, Yun Xiong, and 1 others. 2024.
\newblock Can graph learning improve planning in llm-based agents?
\newblock In \emph{The Thirty-eighth Annual Conference on Neural Information Processing Systems}.

\bibitem[{Xu et~al.(2024)Xu, Cruz, Guevara, Wang, Deshpande, Wang, and Li}]{10.1145/3626772.3661370}
Zhentao Xu, Mark~Jerome Cruz, Matthew Guevara, Tie Wang, Manasi Deshpande, Xiaofeng Wang, and Zheng Li. 2024.
\newblock \href {https://doi.org/10.1145/3626772.3661370} {Retrieval-augmented generation with knowledge graphs for customer service question answering}.
\newblock In \emph{Proceedings of the 47th International ACM SIGIR Conference on Research and Development in Information Retrieval}, SIGIR '24, page 2905–2909, New York, NY, USA. Association for Computing Machinery.

\bibitem[{Ye et~al.(2024)Ye, Zhang, Wang, Xu, and Zhang}]{ye-etal-2024-language}
Ruosong Ye, Caiqi Zhang, Runhui Wang, Shuyuan Xu, and Yongfeng Zhang. 2024.
\newblock \href {https://aclanthology.org/2024.findings-eacl.132/} {Language is all a graph needs}.
\newblock In \emph{Findings of the Association for Computational Linguistics: EACL 2024}, pages 1955--1973, St. Julian{'}s, Malta. Association for Computational Linguistics.

\bibitem[{Zhang(2023)}]{zhang2023graphtoolformerempowerllmsgraph}
Jiawei Zhang. 2023.
\newblock \href {https://arxiv.org/abs/2304.11116} {Graph-toolformer: To empower llms with graph reasoning ability via prompt augmented by chatgpt}.
\newblock \emph{Preprint}, arXiv:2304.11116.

\end{thebibliography}




\end{document}